\documentclass[aps,preprint,prd,epsfig]{revtex4}
\usepackage{amssymb}
\usepackage{amsmath}
\usepackage{graphicx}
\usepackage{fancyhdr}
\newcommand{\be}{\begin{equation}}
\newcommand{\ee}{\end{equation}}
\newcommand{\bea}{\begin{eqnarray}}
\newcommand{\eea}{\end{eqnarray}}
\newcommand{\bes}{\begin{subequations}}
\newcommand{\ees}{\end{subequations}}

\newcommand{\bc}{\begin{center}}
\newcommand{\ec}{\end{center}}
\usepackage{amsmath, amssymb} 
\usepackage{slashed}

\begin{document}

\title{Neutrino mass mechanisms in 3-3-1 models: A short review.}
\author{C. A. de S. Pires}
\email{cpires@fisica.ufpb.br}
\affiliation{{ Departamento de
F\'{\i}sica, Universidade Federal da Para\'\i ba, Caixa Postal 5008, 58051-970,
Jo\~ao Pessoa, PB, Brasil}}

\date{\today}

\begin{abstract}
In this paper we review some  mechanisms that provide light neutrinos in the framework of 3-3-1 gauge models without exotic leptons. In regard to the minimal 3-3-1 model, we call the attention to the fact that the perturbative regime of the model goes until 5 TeV. This requires alternative mechanisms in order to generate light neutrinos. In this review we discuss two mechanisms capable of generating light neutrinos in the framework of the minimal 3-3-1 model. In regard to the 3-3-1 model with right-handed neutrinos, we call the attention to the fact that in it mechanisms that generate light left-handed neutrinos also generate light right-handed neutrinos. Finally,  we  call the attention to the fact that the 3-3-1 model with right-handed neutrinos accommodate naturally the inverse seesaw mechanism.

%
\end{abstract}

\maketitle

\section{Introduction}
\label{sec1}
Models for the electroweak interactions based on the $SU(3)_C \otimes SU(3)_L \otimes U(1)_N$(3-3-1) symmetry were first intensively  explored in the 1970's\cite{earlymodels}.  However, we stress that all those models involved exotic leptons. In general, in these models exotic fermions compose the third component of the fermionic triplet.   Considering the leptonic sector, we could  say that there are two versions of 3-3-1 gauge models. In one version, the third component of the leptonic triplet is a simply charged lepton. The other version involves a neutral lepton as third component of the leptonic triplet. 

In the 1990's it was perceived that the exotic leptons could be replaced by the standard ones. More precisely, in 1992 a 3-3-1 gauge model was proposed where the third component of the leptonic triplet was recognized as the anti-lepton\cite{ppf}. In what follow, in 1993, a second version of the 3-3-1 gauge models was proposed  where  right-handed neutrinos was insightfully proposed as the third component of the leptonic triplet~\cite{footpp}. The first version is nowadays called the minimal version of the 3-3-1 gauge models, minimal 3-3-1 for short,  because its leptonic sector is composed exclusively by the standard leptons, while the second version is called 3-3-1 model with right-handed neutrinos.   In this work we review mechanisms proposed in the literature that generate light neutrinos in the framework  of these two 3-3-1 gauge models.


\section{Neutrino mass mechanisms in the  Minimal 3-3-1 model.}
\label{sec2}

Minimal here means that the  leptonic sector involve  the standard leptons, only. Thus,  each family of leptons  come inside a triplet,
%
\be
f_{l_L}^T = \left (
\nu_l\,, 
e_l\,, 
e^{c}_l
\right )^T_L\sim(1\,,\,3\,,\,0),
\label{leptoniccontent}
\ee
where $l=e,\,\mu ,\, \tau$ and $T$ means transposition. Then, we can immediately conclude that the model may explain electric charged quantization {\it a la} grand unification theories as showed in \cite{ECQ}. 

In addition to the standard gauge bosons, the model has five new ones, namely a new $Z^{\prime}$, two simply charged gauge bosons, $V^{\pm}$ and two doubly charged gauge bosons $U^{\pm \pm}$. We also call the attention to the fact that these four charged gauge bosons carry two units of lepton number each, called bilepton gauge bosons. We stress that the doubly charged gauge bosons constitute genuine and distinguishable signatures of the model\cite{DCS}.

Concerning the scalar sector of minimal 3-3-1 model, the most complex scenario involve three triplets and one scalar sextet.
\begin{eqnarray}
&&\chi^T =
\left (
\chi^- \,, 
\chi^{--} \,, 
\chi^0
\right )^T,\,\,\rho^T =
\left (
\rho^+ \,, 
\rho^0 \,, 
\rho^{++}
\right )^T,\, \nonumber \\
&&\eta^T =
\left (
\eta^0 \,, 
\eta^- \,, 
\eta^+
\right )^T,\,S=
\left (
\begin{array}{lcr}
\sigma^0_1 & \frac{h_2^-}{\sqrt{2}} & \frac{h_1^+}{\sqrt{2}} \\
\frac{h_2^-}{\sqrt{2}} & H_1^{--} & \frac{\sigma_2^0}{\sqrt{2}} \\
\frac{h_1^+}{\sqrt{2}} & \frac{\sigma_2^0}{\sqrt{2}} & H^{++}_2
\end{array}
\right ),
\label{scalarsector} 
\end{eqnarray}
where these scalars have the following transformation properties under the gauge group $SU(3)_C\otimes SU(3)_L\otimes U(1)_N$, $\chi \sim$ ({\bf 1},{\bf 3},-1),  $\rho \sim$ ({\bf 1},{\bf 3},1), $\eta \sim$ ({\bf 1},{\bf 3}, 0) and $S \sim$ ({\bf 1},{\bf 6},0). 
The neutral components of the triplets develop VEV's, $\langle \eta^0\rangle = v_\eta$, $\langle \rho^0\rangle = v_\rho$, $\langle \chi^0\rangle = v_\chi$, so that we have the correct pattern of symmetry breaking, with $v_\chi$ being responsible for the $SU(3)_L\otimes U_N(1)$ breaking to $SU(2)_L\otimes U(1)_Y$ symmetry, while $v_\eta$ and $v_\rho$ combine to break $SU(2)_L\otimes U(1)_Y$ to QED, $U(1)_{e}$.

Considering only one scalar sextet, $S$, the masses of all leptons, including neutrinos, arise from a common set  of Yukawa couplings,
\begin{eqnarray}
G\bar f^C_L S^* f_L.
\label{YukawaS}	
\end{eqnarray}
The sextet $S$ has two neutral components, $\sigma^0_1$ and $\sigma^0_2$, and when both develop nonzero VEV's, $v_{\sigma_1}$  and $v_{\sigma_2}$, neutrinos and charged leptons get their masses given by the following expressions, $m_\nu=Gv_{\sigma_1}\,\,\,\mbox{and}\,\,\, m_l=Gv_{\sigma_2}$.

 Perceive that a unique scalar sextet is not enough to generate the correct masses of all leptons of the model. The reason is simple, notice that the texture of $m_\nu$  is the same as the $m_l$ because both involve the same matrix $G$. Automatically, the rotating mixing matrix, $U$, that diagonalizes $m_\nu$ will diagonalize $m_l$ too. Thus, the lepton mass eigenstates, $\hat{l}$, get related to the symmetry eigenstates, $l$, through the following rotation ,
%
$\hat{\nu}_{l_L} =U_{la} \nu_{a_L}\,\,\,\,\,\,\mbox{and}\,\,\,\,\,\, \hat{e}_{l_L}=U_{la}e_{a_L}\,,$
%
where $a=1,2,3$. Consequently, the charged current will be always diagonal,
%
$	\frac{g}{\sqrt{2}}\bar{\hat{\nu}}_L \gamma^\mu \hat{e}_L W^+_\mu \rightarrow \frac{g}{\sqrt{2}}\bar \nu_L \gamma^\mu e_L W^+_\mu\,,$
%
which goes against the recent atmospheric and solar neutrino oscillation experiments. As an immediate consequence, three scalar triplets and one scalar sextet are not sufficient to explain the masses of all fermions of the model. This observation discards the mechanism developed in Ref. \cite{typeIISS331}. There are, in the literature, many suggestions for solving  this problem  but all of them require the enlargement of the particle content of the model\cite{neutrinos331}.

On the other hand, an interesting predictions of the minimal 3-3-1 model is the one concerning the Weinberg mixing angle, $\theta_W$, which arises from the kinetic term of the gauge bosons and is expressed by the following relation,
\begin{eqnarray}
\frac{g^{\prime}}{g}=\frac{S_W}{\sqrt{1-4S^2_W}}\,,
\label{g-gNrelation}	
\end{eqnarray}
where $S_W=\sin \theta_W$, $g$ is the $SU(3)_L$ coupling and $g^{\prime}$ is the $U(1)_N$ coupling. This relation shows us that a Landau pole exists for the theory and, in order to keep the theory inside the perturbative regime, a bound on this mixing angle is obtained, $S^2_W <0.25$. Translating this in terms of an energy scale, it was pointed out in Ref. \cite{landaupole} that the perturbative regime of the model persists until about few TeV's. In other words, the highest energy scale where the model is perturbatively trustable is about 4-5 TeV and it can certainly be regarded as an effective model so that above that scale the underlying fundamental theory has to be called in. 

In what concern neutrino masses, this is particularly worrying because, once the perturbative scale is around 4-5 TeV, then the effective dimension five operator,
%
\begin{eqnarray}
\frac{h}{\Lambda}(\bar f^C_L \eta^*)(\eta^{\dagger}f_L),
\label{effectiveD-5}
\end{eqnarray}
generates the following mass formula to the neutrinos, 
%
$m_\nu= \frac{h v^2_\eta}{\Lambda}$,
%
then, for $v_\eta \approx 10^2$GeV and $\Lambda=5$TeV, we get $m_\nu= 10h$GeV. In other words, even the effective dimension-5 operator favors heavy neutrinos. So, in what concern neutrinos,  this is the present undesirable status of the minimal 3-3-1 model. This means, too, that all proposals of generating neutrino masses through extension of the minimal 3-3-1 model that did not provide a way out to circumvent the effective dimension-5 operator can not be seeing as a realistic small neutrino mass scenario.

Concerning the problem of small neutrino masses in the minimal 3-3-1 model, as far as I know there are two ways out. The first one make use of $Z_n$ discrete symmetries to avoid dominant effective  operators of low order. For example, for a particular  representation of $Z_3$ we can have effective dimension-11 as dominant effective operators which leads naturally to neutrino at eV scale for $\Lambda =4-5$TeV\cite{trulyminimal}. The second possibility requires the addition of right-handed neutrinos in the singlet form  to the leptonic particle content of the minimal 3-3-1 models and then  combine type I with type II seesaw mechanisms to generate tiny neutrino masses\cite{combinetypeI-II}.

Concerning the first case, on considering the $Z_3$ discrete symmetry with the following representation of the fields,
\begin{eqnarray}
&&	\eta \rightarrow  e^{4i\pi/3}\eta ,\,\rho \rightarrow e^{-4i\pi/3}\rho , \,\chi \rightarrow e^{2i\pi/3} \chi , f_{l_L} \rightarrow e^{2i\pi/3}f_{l_L} ,
\label{Z3symmetry}
\end{eqnarray}
it is amazing that such $Z_3$ symmetry allows, as the first dominant effective operator that generates neutrino mass, exactly a dimension-11 operator,
\begin{eqnarray}
	\frac{h^\nu}{\Lambda^7}  (\bar f^C_L \eta*)( \rho \chi \eta)^2(\eta{\dagger} f_L ).
	\label{D11neutrinos}
\end{eqnarray}
In this equation the term $(\rho \chi \eta)$ is the anti-symmetric singlet combination under $SU_L(3)$.
When $\eta$, $\rho$  and $\chi$ develop their respective VEV's, this operator generates the following  expression for the neutrino masses,
%
$	m_\nu=\frac{h^\nu}{\Lambda^7} v_\eta^4 v_\rho^2 v_\chi^2$.
%
That is a striking result if one observes that even for $\Lambda$ around $v_\chi$, the operator above is suppressed enough to generate small masses for the neutrino. For instance, in being conservative and taking 
$v_\eta \approx 10^2\mbox{GeV}$, $v_\rho \approx 10\mbox{GeV}$, as before,
$v_\chi \approx 1\mbox{TeV}$ and $\Lambda =5\mbox{TeV}$,
we obtain a prediction for the neutrino masses,	$m_\nu \approx 0.1 h^\nu \mbox{eV}$, meaning that
a $\Lambda$ around few TeV's is perfectly compatible with neutrino masses at the sub-eV scale, which represents an astonishing achievement for the model. 

Concerning the second case, it was showed in Ref. \cite{twotriplets} that the minimal 3-3-1 model may be implemented with two scalar triplets, only, namely, $\chi$ and $\rho$. This is the reduced version of the 3-3-1 model\cite{rho}. In the latter charged leptons and some quarks gain masses through effective dimension-5 operators. This is quite possible because $\Lambda =4-5$TeV. In order to obtain small neutrino masses in this version, we add right-handed neutrinos in the singlet form and  a third scalar triplet $\phi=(\phi^{0}\,\,\,\, \phi_1^{-}\,\,\, \phi_2^{+})^T \sim (1,3,0)$ to the other two $\chi$ and $\rho$ with the main supposition being that   $\phi$ is heavier than the other with its mass being around $\Lambda$\cite{combinetypeI-II}.

In this point, it is imperative to impose a $Z_4$ symmetry with the following fields transforming as,
\begin{eqnarray}
&&\rho \rightarrow w^3 \rho\,,\,\chi \rightarrow w^3 \chi\,,\,\phi \rightarrow w^2 \phi\,\,\,,\,\nu_{a_R} \rightarrow w^0 \nu_{a_R}\, , \,f_{a_L} \rightarrow w^2 f_{a_L},
\label{Z4}
\end{eqnarray}
where $w=e^{i\frac{\pi}{2}}$.

In view of this, neutrino mass terms can only arise from the Yukawa interactions below,
\begin{equation}\label{yukawaterms}
{\cal  L} =Y \overline{f}_{L}\phi\nu_{R} + \frac{M}{2}\overline{\nu}_{R}^c\nu_{R} + H.c.,
\end{equation}
where $Y$ and $M$ are  $3\times3$ matrices. 

All neutral fields  $\rho^0$,  $\chi^{ 0}$ , $\phi^0$  are assumed to develop VEV according to
\begin{eqnarray}
\langle\rho^0\rangle=\frac{v_{\rho}}{\sqrt2},\label{veta}\,\,\,\,\,\,
 \langle\chi^{\prime }\rangle=\frac{v_{\chi^{\prime }}}{\sqrt2}\,\,\,\,\,
\langle\phi^0\rangle=\frac{v_{\phi}}{\sqrt2},\,\,\,\,\,\,
\label{vrho}
\end{eqnarray}

When  $\phi$ develop VEV different from zero, neutrinos develop  Dirac   mass terms and we have,
\begin{equation}\label{yukawaterms}
{\cal  L}_{\mbox{mass}} =\frac{Yv_\phi}{\sqrt{2}} \overline{\nu}_{L}\nu_{R} +\frac{M}{2} \overline{\nu}_{R}^c\nu_{R} + H.c.
\end{equation}

On considering the basis $\nu=(\nu_{L}, \nu_{R}^{C})$, we can write the mass terms above in the form,
\begin{equation}\label{neutrinomassterms}
{\cal L}_{\mbox{mass}} =\frac{1}{2}\bar \nu^C M_\nu \, \nu + H.c.,
\end{equation}
where,
\begin{equation}M_{\nu}= \left(\begin{array}{cc}
0 & M_D\\
M^T_D&  M\\
\end{array}\right),
\end{equation}
with $M_D=\frac{Yv_{\phi}}{\sqrt{2}}$ and $M$. When all values of $M$  are larger than all elements of  $ M_D$, we obtain, after diagonalizing this mass matrix, the following expressions for the left and right-handed neutrino masses
\begin{equation} \label{matriz}
m_{\nu_{l}}  \approx M^T_D M^{-1} M_D, \quad m_{\nu_{R}} \approx M.
\end{equation}

This is the canonical type I seesaw mechanism\cite{typeI}. As usual we consider  $M$ diagonal and degenerate.

In regard to the left-handed(LH) neutrinos, according to Eq. (\ref{matriz}) the order of magnitude of their masses  is,
\begin{equation} \label{seesawI}
m_{\nu_{l}}  \approx \frac{v^{2}_{\phi}}{M}.
\end{equation}
As it is known, the maximum value $M$ can acquire  is around $5$ TeV\cite{landaupole}. Thus for we have LH neutrinos at the eV scale, we need $v_\phi \approx10^{-2} $GeV. There is no lower bound on this parameter.   The upper bound, $v^2_\phi + v^2_\rho=(246)^2$GeV, arises because  $v_\rho$ as well as $v_\phi$ both contribute to the mass of the standard charged gauge boson $W^{\pm}$. Thus there is no problem in taking $v_\phi$ as small as we wish since we have $v_\rho \approx 246$GeV.

A suppressed $v_\phi$ can be obtained  through a kind of type II seesaw mechanism\cite{typeII} as developed in \cite{grimus}. For this we consider the most complete potential gauge invariant under 3-3-1  and that conserve $Z_4$,
\begin{eqnarray} \label{potential}
V(\phi, \rho, \chi, \sigma) & = & \mu_{\phi}^{2}\phi^{\dagger}\phi + \mu_{\rho}^{2}\rho^{\dagger}\rho + \mu_{\chi}^{2}\chi^{\dagger}\chi \nonumber\\
&+& \lambda_{1}(\phi^{\dagger}\phi)^{2} + \lambda_{2}(\rho^{\dagger}\rho)^{2} + \lambda_{3}({\chi}^{\dagger}\chi)^{2} + \nonumber\\
&+& \lambda_{4}(\phi^{\dagger}\phi)(\rho^{\dagger}\rho) + \lambda_{5}(\phi^{\dagger}\phi)(\chi^{\dagger}\chi) + \lambda_{6}(\rho^{\dagger}\rho)(\chi^{\dagger}\chi) \nonumber\\
&+& \lambda_{7}(\rho^{\dagger}\phi)(\phi^{\dagger}\rho) + \lambda_{8}(\chi^{\dagger}\phi)(\phi^{\dagger}\chi) + \lambda_{9}(\rho^{\dagger}\chi)(\chi^{\dagger}\rho)\nonumber\\
&-& \frac{f}{\sqrt{2}} \epsilon^{ijk} \phi_{i} \rho_{j}\chi_{k} +H.c.,
\end{eqnarray}
where $f$ is a free parameter with dimension of mass.

From this potential, we obtain the following minimum condition to  $\phi$ develop vev,
\begin{eqnarray}
v_{\phi} (\mu_{\phi}^{2} + \lambda_{1}v_{\phi}^{2} + \frac{1}{2}\lambda_{4}v_{\rho}^{2} + \frac{1}{2}\lambda_{5}v_{\chi}^{2})  - \frac{1}{2}f v_{\rho}v_{\chi}  = 0,
\label{vevconstraints}
\end{eqnarray}

The highest energy scale  the model support is around 4-5TeV. We assume that the scalar $\phi$  belong to this energy scale. This means that  $\mu_\phi \approx M$. Note that  $\mu_\phi$ is dominant in the parenthesis in the first relation of the Eq.  ({\ref{vevconstraints}). As a result, we obtain from it,
\begin{equation} \label{seesawII}
v_{\phi} = \dfrac{ f v_{\rho} v_{\chi}}{2 M^2}.
\end{equation}

On substituting this expression for $v_\phi$ in Eq. (\ref{matriz}), we obtain the following expression to the LH neutrino mass,
\begin{equation}
m_\nu=\frac{\sqrt{2}}{8}Y^T Y\frac{ f^2 v^2_\rho v^2_\chi }{M^5}.
\label{finalmassmatrix}
\end{equation}
This is the main point of our work.  According to this expression neutrino masses get suppressed by high scale $M^5$ in its denominator. This allows we have neutrino masses at eV scale for $M$ around few TeV's. For we see this note that  for typical values $f=1$GeV, $v_\rho=246$GeV, $v_\chi=10^3$GeV and $M=5\times10^3$GeV, we obtain,
\begin{equation}
m_\nu=3.4 Y^T Y\mbox{eV},
\label{magnitude}
\end{equation}
which falls exactly at the eV scale. Whatever comes to be the texture of the neutrino masses, it can be   obtained by choosing an adequate values for the elements of the  matrix $Y$.

\section{Neutrino mass mechanisms in the  3-3-1 model with right-handed neutrinos.}
\label{sec3}
The model we consider is the 3-3-1 model with right-handed neutrinos~\cite{footpp}. It is one of the possible models allowed by the $SU(3)_C \otimes SU(3)_L \otimes U(1)$  gauge symmetry where  leptons are distributed in the following representation content,
\begin{eqnarray}
L_{a} = \left (
\begin{array}{c}
\nu_{aL} \\
e_{aL} \\
(\nu_{aR})^{c}
\end{array}
\right )\sim(1\,,\,3\,,\,-1/3)\,,\,\,\,e_{aR}\,\sim(1,1,-1),
 \end{eqnarray}
where $a = 1,\,2,\, 3$ refers to the three generations. After the spontaneous breaking of the 3-3-1 symmetry to the standard one, the leptonic triplets above splits into the standard leptonic doublet $L_a=\left( \nu_{aL} \,\,,\,\, e_{aL}\right)^T$ plus the singlet $( \nu_{aR})^C$. Thus this model recovers the standard model with right-handed neutrinos.

We remember here that it is not a trivial task to generate light right-handed neutrinos in any simple extension of the standard model. However, in the 3-3-1 model in question, right-handed neutrinos can naturally obtain small masses through effective dimension-five operators. This is due, in part, to the fact that, in the model, the right-handed neutrinos compose, with the left-handed ones,  the same leptonic triplet $L$. As we will see here, it is this remarkable feature that turns feasible  the raise of light right-handed neutrinos.

The gauge sector of the model is composed by the standard gauge bosons  and others five  gauge bosons called  $V^{\pm}$, $U^0$, $U^{0 \dagger}$ and $Z^{\prime}$. 
Also, the model possesses three scalar triplets, two of them transforming as,  $\eta \sim ({\bf 1}\,,\,{\bf 3}\,,\,-1/3)$ and $\chi \sim ({\bf 1}\,,\,{\bf 3}\,,\,-1/3)$
and the other as, $\rho \sim ({\bf 1}\,,\,{\bf 3}\,,\,2/3)$, with the following vacuum structure,
\begin{eqnarray}
\langle \eta \rangle_0 = \left (
\begin{array}{c}
 \frac{v_\eta}{\sqrt{2}} \\
 0\\
 0 
\end{array}
\right ),\,\langle \rho \rangle_0 =\left (
\begin{array}{c}
0 \\
\frac{v_\rho}{\sqrt{2}} \\
0
\end{array}
\right ) ,\, \langle \chi \rangle_0 = \left (
\begin{array}{c}
0 \\
0 \\
\frac{v_{\chi^{\prime}}}{\sqrt{2}}
\end{array}
\right )\,. \label{VEVstructure} 
\end{eqnarray}
These scalars are sufficient to engender spontaneous symmetry breaking and generate the correct masses for all massive particles.

In order to have the minimal model, we assume the following discrete symmetry transformation over the full Lagrangian,
\begin{eqnarray}
&&\left( \chi,\,\eta,\,\rho,e_{aR},\right) \rightarrow -\left( \chi,\,\eta,\,\rho,e_{aR},\,\right),
\nonumber \\
\label{discretesymmetryI}
\end{eqnarray}
where $a=1,2,3$  and $i=1,2$. This discrete symmetry helps in avoiding  unwanted Dirac mass term for the neutrinos~\cite{footpp} and implies a realistic minimal  potential~\cite{pal}. 
  
In the original version of the model, all massive particles gain masses, except neutrinos.  Thus we construct all possible effective dimension-five operators in the 3-3-1 model with right-handed neutrinos that lead to  neutrino masses\cite{naturallylightRH}. The first one involves the leptonic  triplets $L$ and the scalar triplet  $\eta$. With these triplets we can form the following  effective dimension-five operator,
\begin{eqnarray}
{\mathcal L}_{M_L}&&=	\frac{f_{ab}}{\Lambda}\left( \overline{L^C_a} \eta^* \right)\left( \eta^{\dagger}L_{b} \right)+\mbox{H.c}.
	\label{5dL}
\end{eqnarray}	
According to this operator, when $\eta^0$ develops a VEV, $v_\eta$, the left-handed neutrinos develop  Majorana mass terms with the  form,
\begin{eqnarray}
(M_{L})_{ab}=\frac{f_{ab}v^2_\eta}{\Lambda}.
	\label{nuL}
\end{eqnarray}

Due to the fact that right-handed neutrinos are not singlets in the model in question, a second  effective dimension-five operator generating neutrino masses is possible. It is constructed with the scalar triplet $\chi$ and the leptonic triplet $L$,
\begin{eqnarray}
{\mathcal L}_{M_R}&&=	\frac{h_{ab}}{\Lambda}\left( \overline{L_{a}^C} \chi^* \right)\left( \chi^{\dagger}L_{b} \right)+\mbox{H.c}.
	\label{5dR}
\end{eqnarray}	
When $\chi^{\prime 0}$ develops a VEV, $v_{\chi^{\prime}}$, this effective operator  provides  Majorana masses for the right-handed neutrinos,
\begin{eqnarray}
(M_{R})_{ab}\overline{(\nu_{aR})^C} \nu_{bR},\,\,\,\,\mbox{with} \,\,\,\,(M_{R})_{ab}=\frac{h_{ab}v^2_{\chi^{\prime}}}{\Lambda}
	\label{nuR}
\end{eqnarray}

Thus, we have Majorana  mass terms for the neutrinos  both having the same origin, i.e.,  effective dimension-five operators\cite{LRversion}. 
 At this point, two comments are in order. First, as the VEV $v_{\chi^{\prime}}$ is responsible for the breaking of the 3-3-1 symmetry to the standard symmetry, and that  $v_\eta$ contributes to the spontaneous breaking of the standard symmetry, thus it is natural to expect that $v_{\chi^{\prime}}> v_\eta$, which implies $M_R>M_L$.  For example. Taking $\Lambda=10^{12}$ GeV, $v_\eta=40$GeV and $v_{\chi^{\prime}}=5$TeV, we obtain $(M_{L})_{ab}=1.6 f_{ab} $eV and $(M_{R})_{ab}=25 h_{ab} $keV. Thus, in the 3-3-1 model with right-handed neutrinos, we have left-handed neutinos with mass at eV scale and right-handed neutrinos with mass at keV scale. Those right-handed neutrinos are natural candidates for warm dark matter.

These effective operators may be realized through the addition of at least one scalar sextet\cite{realization},
\begin{equation}
S=
\dfrac{1}{\sqrt{2}}
\begin{pmatrix}
\Delta^{0} & \Delta^{-} & \Phi^{0} \\
\Delta^{-} & \Delta^{--} & \Phi^{-} \\
\Phi^{0} & \Phi^{-} & \sigma^{0}
\end{pmatrix}
\sim (1,6,\frac{-2}{3}).
\end{equation}
With the leptonic triplet $L$  and the sextet $S$ we form the Yukawa coupling $\bar L SL^C $. The sextet $S$ has three neutral scalars that may develop VEV. The VEV's of the neutral scalars  $\Delta^0$ and  $\sigma^0$ will lead to  Majorana mass terms for both the left-handed and right-handed neutrinos respectively, while the VEV of the neutral scalar  $\Phi^0$  will generate a Dirac mass term which mix the left-handed neutrinos with the  right-handed ones. In order to implement the type II seesaw mechanism the  Dirac mass terms must be avoided. For this we assume that  $\Phi^0$ does not develop VEV and impose the set of discrete symmetries $(\chi,\rho, e_{aR}) \rightarrow -(\chi,\rho, e_{aR})$. The discrete symmetry also helps in avoiding flavor changing neutral currents involving quarks and scalars and is important to obtain a simple potential. In regard to the Dirac mass terms, we think important to emphasize that, in avoiding them,  the left-handed neutrinos get decoupled of the right-handed ones. Thus, in this case, we could say that our right-handed neutrinos are, in fact, completely steriles. For this reason, from now on we refer to these neutrinos as the sterile neutrinos.

The mechanism arises in the potential of the model and is communicated to the neutrinos through the Yukawa interaction $\bar L L^C S$. The essence of the mechanism is that lepton number is violated explicitly through some terms in the potential. For this it is necessary to know the lepton number distribution of the scalars: $L(\eta^{\prime 0}\,,\,\sigma^0 \,,\,\rho^{\prime +})=-2$  and  $L(\chi^0\,,\, \chi^-\,,\,\Delta^0\,,\,\Delta^-\Delta^{--})=2$.  When $\Delta^0$ and $\sigma^0$ develop VEV's, automatically the left-handed neutrinos($\nu_L$)  and the sterile ones($\nu_R$) both  develop  Majorana mass terms.  The masses of $\nu_L$ and  of $\nu_R$ get proportional to $v_\Delta$  and  $v_\sigma$, respectively. The role of the type II seesaw mechanism  is to furnish tiny values for $v_\Delta$  and $v_\sigma$.

The most complete part of the potential that obeys the discrete symmetry discussed above and conserves lepton number is composed by the following terms, 
\begin{eqnarray}
V & = & \mu_{\chi}^{2}\chi^{2}+\mu_{\eta}^{2}\eta^{2}+\mu_{\rho}^{2}\rho^{2}+\lambda_{1}\chi^{4}+\lambda_{2}\eta^{4}+\lambda_{3}\rho^{4}+\lambda_{4}(\chi^{\dagger}\chi)(\eta^{\dagger}\eta) \nonumber\\
  & + & \lambda_{5}(\chi^{\dagger}\chi)(\rho^{\dagger}\rho)+\lambda_{6}(\eta^{\dagger}\eta)(\rho^{\dagger}\rho)+\lambda_{7}(\chi^{\dagger}\eta)(\eta^{\dagger}\chi)+\lambda_{8}(\chi^{\dagger}\rho)(\rho^{\dagger}\chi) \nonumber\\
  & + & \lambda_{9}(\eta^{\dagger}\rho)(\rho^{\dagger}\eta)+(\frac{f}{\sqrt{2}}\epsilon^{ijk}\eta_{i}\rho_{j}\chi_{k}+H.C)+\mu_{S}^{2}Tr(S^{\dagger}S) \nonumber\\
  & + & \lambda_{10}Tr(S^{\dagger}S)^{2}+\lambda_{11}[Tr(S^{\dagger}S)]^{2}+(\lambda_{12}\eta^{\dagger}\eta+\lambda_{13}\rho^{\dagger}\rho+\lambda_{14}\chi^{\dagger}\chi)Tr(S^{\dagger}S) \nonumber\\
  & + & \lambda_{15}(\epsilon^{ijk}\epsilon^{lmn}\rho_{n}\rho_{k}S_{li}S_{mj}+H.C)+\lambda_{16}(\chi^{\dagger}S)(S^{\dagger}\chi)+\lambda_{17}(\eta^{\dagger}S)(S^{\dagger}\eta) \nonumber\\
  & + & \lambda_{18}(\rho^{\dagger}S)(S^{\dagger}\rho),
\end{eqnarray}
while the other part  that violates explicitly the lepton number is composed by these terms,  
\begin{eqnarray}
V^{\prime} & = &
\lambda_{19}(\eta^{\dagger}\chi)(\eta^{\dagger}\chi)+(\frac{\lambda_{20}}{\sqrt{2}}\epsilon^{ijk}\eta_{m}^{*}S_{mi}\chi_{j}\rho_{k}+H.C)+(\frac{\lambda_{21}}{\sqrt{2}}\epsilon^{ijk}\chi_{m}^{*}S_{mi}\eta_{j}\rho_{k}+H.C)\nonumber\\
        & - & M_{1}\eta^{T}S^{\dagger}\eta-M_{2}\chi^{T}S^{\dagger}\chi.
\end{eqnarray}
We think we have presented all the aspects of the model that are relevant to the implementation of the type II seesaw mechanism, which we do next. 
From the Yukawa interaction,
\begin{eqnarray}
	{\cal L}^Y_\nu = G_{ab}\bar L_{aL} S L^C _{bL} + \mbox{H.c},
	\label{yukawaforneutrinos}
\end{eqnarray}
when $\Delta^0$  and $\sigma^0$ both develop VEV,  the left-handed and the sterile neutrinos develop the following mass terms, 
\begin{eqnarray}
	{\cal L}^Y_\nu = G_{ab}v_\Delta\bar \nu_{aL}^C \nu_{bL}  + G_{ab}v_\sigma\bar \nu_{aR}^C \nu_{bR} .
	\label{preliminarymassmatrix}
\end{eqnarray}
We emphasize here that the mass terms of both neutrinos have as common origin the Yukawa interaction in Eq. (\ref{yukawaforneutrinos}). In practical terms this means that the same Yukawa couplings  $G_{ab}$ are common for the left-handed and sterile neutrino masses. That's a very interesting result because when the masses of the left-handed neutrinos get measured directly, automatically the masses of the sterile neutrinos will be predicted.

The role of the type II seesaw mechanism here is to provide tiny values for  $v_\Delta$  and $v_\sigma$.  This is achieved from the minimum condition of the  potential $V+V^{\prime}$. For this it is necessary to select which of the eight neutral scalars of the model  develop VEV. We already discussed why  $\Phi^0$ is not allowed here to develop VEV. In  the original version of the model only $\eta^0$, $\rho^0$  and $\chi^{\prime o}$ were allowed to develop VEV. The reason for this is to avoid flavor changing neutral currents involving  quarks and scalars. Thus we have that the simplest case is when only $\chi^{\prime 0}\,,\, \rho^{0}\,,\, \eta^{0}\,,\, \Delta^{0}\,,\,\sigma^{0} $  develop VEV. We assume this and shift these fields in the following way,
\begin{equation}
\chi'^{0}, \rho^{0}, \eta^{0}, \Delta^{0}, \sigma^{0} \rightarrow \dfrac{1}{\sqrt{2}}(v_{\chi', \rho, \eta, \Delta, \sigma}+R_{\chi', \rho, \eta, \Delta, \sigma}+iI_{\chi', \rho, \eta, \Delta, \sigma})
\label{shift}.
\end{equation}

Considering the shift of the neutral scalars in Eq. (\ref{shift}), the composite potential, $V+V^{\prime}$, provides the following set of minimum conditions,

\begin{eqnarray}
\mu_{\rho}^{2}&&+\lambda_{3}v_{\rho}^{2}+\frac{\lambda_{5}}{2}v_{\chi'}^{2}+\frac{\lambda_{6}}{2}v_{\eta}^{2}+\frac{fv_{\eta}v_{\chi'}}{2v_{\rho}}+\frac{\lambda_{13}}{4}(v_{\sigma}^{2}+v_{\Delta}^{2})+\nonumber\\
  &   & \lambda_{15}v_{\Delta}v_{\sigma}-\frac{\lambda_{20}}{4}(\frac{v_{\eta}v_{\Delta}v_{\chi'}}{v_{\rho}})+\frac{\lambda_{21}}{4}(\frac{v_{\chi'}v_{\eta}v_{\sigma}}{v_{\rho}})=0,
  \nonumber \\
  \mu_{\eta}^{2}&&+\lambda_{2}v_{\eta}^{2}+\frac{\lambda_{4}}{2}v_{\chi'}^{2}+\frac{\lambda_{6}}{2}v_{\rho}^{2}+\frac{fv_{\chi'}v_{\rho}}{2v_{\eta}}+\frac{\lambda_{12}}{4}(v_{\sigma}^{2}+v_{\Delta}^{2})+\nonumber\\
  &   & \frac{\lambda_{17}}{4}v_{\Delta}^{2}-M_{1}v_{\Delta}-\frac{\lambda_{20}}{4}(\frac{v_{\Delta}v_{\rho}v_{\chi'}}{v_{\eta}})+\frac{\lambda_{21}}{4}(\frac{v_{\chi'}v_{\rho}v_{\sigma}}{v_{\eta}})=0,\nonumber \\
  \mu_{\chi}^{2}&&+\lambda_{1}v_{\chi'}^{2}+\frac{\lambda_{4}}{2}v_{\eta}^{2}+\frac{\lambda_{5}}{2}v_{\rho}^{2}+\frac{fv_{\eta}v_{\rho}}{2v_{\chi'}}+\frac{\lambda_{14}}{4}(v_{\Delta}^{2}+v_{\sigma}^{2})+\nonumber\\
  &   & \frac{\lambda_{16}}{4}v_{\sigma}^{2}-M_{2}v_{\sigma}-\frac{\lambda_{20}}{4}(\frac{v_{\eta}v_{\rho}v_{\Delta}}{v_{\chi'}})+\frac{\lambda_{21}}{4}(\frac{v_{\eta}v_{\rho}v_{\sigma}}{v_{\chi'}})=0,\nonumber \\
  \mu_{S}^{2}&&+\frac{\lambda_{10}}{2}v_{\Delta}^{2}+\frac{\lambda_{11}}{2}(v_{\sigma}^{2}+v_{\Delta}^{2})+\dfrac{\lambda_{12}}{2}v_{\eta}^{2}+\dfrac{\lambda_{13}}{2}v_{\rho}^{2}+\dfrac{\lambda_{14}}{2}v_{\chi'}^{2}+\lambda_{15}\frac{v_{\rho}^{2}v_{\sigma}}{v_{\Delta}}+
  \nonumber\\
 && \dfrac{\lambda_{17}}{2}v_{\eta}^{2}-\frac{\lambda_{20}}{2}(\frac{v_{\eta}v_{\rho}v_{\chi'}}{v_{\Delta}})-M_{1}\frac{v_{\eta}^{2}}{v_{\Delta}}=0,
  \nonumber \\ \mu_{S}^{2}&&+\frac{\lambda_{10}}{2}v_{\sigma}^{2}+\frac{\lambda_{11}}{2}(v_{\sigma}^{2}+v_{\Delta}^{2})+\dfrac{\lambda_{12}}{2}v_{\eta}^{2}+\dfrac{\lambda_{13}}{2}v_{\rho}^{2}+\dfrac{\lambda_{14}}{2}v_{\chi'}^{2}+\lambda_{15}\frac{v_{\rho}^{2}v_{\Delta}}{v_{\sigma}}+\nonumber\\
  &   & \dfrac{\lambda_{16}}{2}v_{\chi'}^{2}-M_{2}\frac{v_{\chi'}^{2}}{v_{\sigma}}+\frac{\lambda_{21}}{2}(\frac{v_{\eta}v_{\rho}v_{\chi'}}{v_{\sigma}})=0.
  \label{minimum}
\end{eqnarray}

In any conventional seesaw mechanism,  the masses of the particles inherent of the mechanism and the energy scale associated to the  violation of the lepton number both must  lie in the GUT regime. Bringing this to our mechanism, we have that the masses of the scalars that compose the sextet, $\mu_S$, and the energy scale $M_1$  and $M_2$ that appear in terms that violated explicitly the lepton number both must lie in the GUT energy regime which is around $10^{12}$GeV. For sake of simplicity, here we assume  $\mu_S\approx M_1 \approx M_2=M$.

As   $v_{\sigma\,,\,\Delta} << v_{\chi^{\prime}, \rho, \eta}$  and $v_{\chi^{\prime}, \rho, \eta} << M$, we see that the fourth expression in  Eq. (\ref{minimum}) provides $v_\Delta = \frac{v^2_\eta}{M}$  while the last one provides $v_\sigma = \frac{v^2_{\chi^{\prime}}}{M}$.   On substituting these expressions for the VEV's $v_\Delta $  and  $v_\sigma$ in Eq. (\ref{preliminarymassmatrix}), we obtain,
\begin{eqnarray}
	m_{\nu_L}=G\frac{v^2_{\eta}}{M}\,\,\,\,\, , \,\,\,\,\,m_{\nu_R}= G\frac{v^2_{\chi^{\prime}}}{M}.
	\label{finalmassmatrix}
\end{eqnarray}
Notice that, the higher the $M$, the smaller the masses of $\nu_L$  and $\nu_R$. This is our type II seesaw mechanism where the masses of the neutrinos are suppressed by the high energy scale $M$. To go further, and make some predictions, there are no other way unless take fine-tunning of the free parameters involved in the mechanism. However, note that, apart from the Yukawa coupling $G_{ab}$, we have that $\frac{m_{\nu_R}}{m_{\nu_L}}=\frac{v^2_{\chi^{\prime}}}{v^2_{\eta}}$. For typical values of both VEV's, for example, $v_{\eta}=10^2$GeV  and $v_{\chi^{\prime}}=10^4$GeV we obtain $\frac{m_{\nu_R}}{m_{\nu_L}}=10^4$. Thus, for left-handed neutrinos of mass of order of $10^{-1}$eV  we may have sterile neutrinos of mass of few KeV. This is an encouraging result because sterile neutrinos with mass in this range  is a viable warm dark matter candidate\cite{dodelson}.
\section{ Simple Realization of the Inverse Seesaw Mechanism}
\label{sec4}
The inverse seesaw(ISS) mechanism is a phenomenological small mass generation mechanism  for the neutrinos since it works at TeV scale and then may be probed with the present colliders as LHC\cite{ISSgeneral}.

For the implementation of the ISS mechanism in the 3-3-1 model with right-handed neutrinos we need to  enlarge the leptonic sector of the model by adding three singlet neutral fermions\cite{ISS331}
\begin{eqnarray}
& & f_{aL} = \left (
\begin{array}{l}
\nu_{aL} \,\, e_{aL} \,\, \nu^{C}_{aL}
\end{array}
\right )^T\sim(3\,,\,-1/3)\,,\nonumber\\
& & e_{a_R}\,\sim(1,-1)\,,\,\,\,\,\,\,\,\,\,N_{a_L}\sim(1,0),
\end{eqnarray}
where $a=1,2,3$\cite{moreonISS331}.

The relevant Yukawa Lagrangian for the lepton sector that yields the ISS mechanism for the neutrinos is composed by the following summation of terms,
\begin{equation}
{\cal L}^Y_{\mbox{ISS}}=G_{ab}\epsilon_{ijk}\bar{L^C_{a_i}}\rho^*_j L_{b_k} + G^{\prime}_{ab}\bar L_a \chi (N_{b_L})^C + \frac{1}{2} \bar N^C_L \mu N_L + H.c.
\label{yukawainteractions}
\end{equation}

With the set of VEV's of Eq. (\ref{VEVstructure} ), and considering the basis  $S_L =(\nu_L\,,\, \nu^{C}_L\,,\,N_L)$, the mass terms above can be cast in the following manner,
\begin{equation}
{\cal L}_{\mbox{mass}}=\frac{1}{2} \bar S^C_L M_\nu  S_L + H.c.,
\label{massterm}
\end{equation}
with the mass matrix $M_\nu$ having the texture,
\begin{equation}
M_\nu=
\begin{pmatrix}
0 & m^T_D & 0 \\
m_D& 0 & M^T\\
0 & M & \mu
\end{pmatrix}.
\label{massmatrix331}
\end{equation}
where the $3\times3$ matrices are defined as
\begin{eqnarray}
& & M_{ab}=G^{\prime}_{ab}\frac{v_{\chi^{\prime}}}{\sqrt2}\label{mM}\\
& & m_{Dab}=G_{ab}\frac{v_\rho}{\sqrt2}
\label{mmatrix3x3}
\end{eqnarray}
with  $M_{ab}$ and $m_{D_{ab}}$ being Dirac mass matrices, with this last one being anti-symmetric. The mass matrix in Eq.~(\ref{massmatrix331}) is characteristic of the ISS mechanism. We would like to call the attention to the fact  that the two energy scales related with the model's gauge symmetry breakdown appear in the mass matrix. Namely,  $v_{\chi^{\prime}}$ in $M_{ab}$ is connected with $SU(3)_L \otimes U(1)_N/SU(2)_L\otimes U(1)_Y$ and could be expected to be at the TeV scale leading to observable effects at the LHC, while $v_{\rho}$ in $m_{Dab}$ is connected with the electroweak standard model symmetry breakdown scale. The third  scale of energy, $\mu$, is a energy scale  characteristic of the ISS mechanism and  lie at KeV scale.

The mass matrix above is the typical one of the ISS mechanism whose diagonalization  leads to three light Majorana neutrinos given by, 
\begin{eqnarray}
m_{light} = m^T_D M^{-1}\mu (M^T)^{-1}m_D,
\label{mlf}
\end{eqnarray}
and six heavy new neutrinos with masses around TeV scale.

Returning to $m_{light}$,  on substituting  $m_D=Gv_\rho$, $M=G^{\prime} v_{\chi^{\prime}}$, we obtain
\begin{equation}
m_{light} =\left( G^T (G^{ \prime T})^{-1}\mu (G^{ \prime})^{ -1} G\right)\frac{v^2_\rho }{v^2_{\chi^{\prime}}}.
\label{inverseseesaw331}
\end{equation}

Remember that  $G$ is an anti-symmetric matrix, implying that one eigenvalue of the neutrino mass matrix in Eq.~(\ref{inverseseesaw331}) is null. For $v_\rho \simeq 10^2$GeV,  $v_{\chi^{\prime}}\simeq 10^3$ GeV and $\mu \simeq 10^{-7}$ GeV, we obtain two light neutrinos with masses around eV.

\section{Conclusions}
\label{sec5}
The explanation of the lightness of the neutrino masses is a challenge for any gauge theory. This is also the case for models based on the 3-3-1 gauge symmetry. This is particularly true for the case of the minimal 3-3-1 model where perturbative regime is trustable until 4-5 TeV. We know that the most easy and elegant way of generating small neutrino masses is supposing they are consequence of physics at GUT scale. This idea is realized through seesaw mechanisms. The difficulty of the minimal 3-3-1 model  to generate small neutrino masses is that there effective dimension-5 operators exist and are suppressed by an energy scale of order of 5 TeV\cite{rho}. Thus, as a first guess we  looked for discrete symmetries that avoid dangerous effective operators of low order. This was done in Ref. \cite{trulyminimal} by imposing a $Z_3$ symmetry with a special representation. Other possibility, which do not need to worry with effective dimension-5 operators, was presented in Ref. \cite{combinetypeI-II} where a combination of type I with type II seesaw mechanisms where used to obtain theoretically consistent model of small neutrino masses.

In regard to the 3-3-1 with right-handed neutrinos, differently from the minimal 3-3-1 case,  effective dimension-five operators are sufficient to generate light neutrino masses. However, as the model  has the right-handed neutrinos composing, together with the left-handed neutrinos, the same leptonic triplet, then, as nice result, the model provides left-handed and right-handed light neutrinos. The  effective dimension-5 operator may be realized  through a kind of type II seesaw mechanism implemented by a sextet of scalars belonging to the GUT scale.  The interesting point here is that the left-handed as well as the right-handed neutrino both develop light masses with the right-handed ones having masses  at KeV scale In view of this, the lightest of them may be stable and constitute some portion of the dark matter of the universe in the form warm dark matter. For completeness reasons, we would like to call the attention to the fact that, as far as I know,  minimal versions of the 3-3-1 models do not  contain in their particle spectrum  any candidate that fulfill  the condition required to be  a dark matter candidate, while 3-3-1 models with right-handed neutrinos present dark matter candidate in the fermion and scalar forms\cite{DM331}. 

Another positive aspect concerning the 3-3-1 model with right-handed neutrinos is that it realizes naturally the inverse seesaw mechanism. Thus, we have a model with a seesaw mechanism that may be probed at the LHC and future colliders once such a mechanism works at TeV scale. Thus, in what concern neutrino masses, the 3-3-1 model with right-handed neutrinos seems to be more attractive than the minimal 3-3-1 model. 

In what concern LHC and 3-3-1 signature, many things remain to be done.  Apart from some works considering the Higgs phenomenology \cite{Higgspheno}, we have constraint on the $Z^{\prime}$ mass being abobe 2TeV\cite{zprime} and some phenomenological aspects considering charged scalar production\cite{diego}

\acknowledgments
I would like to thank my collaborators  P. S. Rodrigues da Silva, A. G. Dias and F. Queiroz. This work was supported by Conselho Nacional de Pesquisa e Desenvolvimento Cient\'{i}fico- CNPq. 


\begin{thebibliography}{99}
\bibitem{earlymodels}
For a list of references although incomplete, see: J. Schechter and Y. Ueda, Phys. Rev.  D {\bf 8} (1973), 484; J.  G. Segr\`e and J. Weyers, Phys. Lett.  B {\bf 65} (1976), 243; H. Fritzch and P. Minkowski, Phys. Lett. 
B {\bf 63}(1976), 204; B. W. Lee and S. Weinberg, Phys. Rev.  Lett. {\bf 38} (1977), 1237; 
B. W. Lee and R. E. Shrock, Phys. Rev. D {\bf 17} (1978), 2410;  P. 
Langacker, G. Segr\`e and M. Golshani, Phys. Rev.  D {\bf 17}(1978), 1402; M. Singer, J. W. 
F. Valle and J. Schechter, Phys. Rev.  D {\bf 22} (1980), 738.
%
\bibitem{ppf}
F. Pisano and V. Pleitez, Phys. Rev.  D {\bf 46 }(1992), 410; 
 P. H. 
Frampton, Phys. Rev. Lett. {\bf 69} (1992), 2889.
%
\bibitem{footpp}
R. Foot, H. N. Long and T. A. Tran, Phys. Rev. D{\bf 50}
(1994), R34; J. C. Montero, F. Pisano and V. Pleitez, Phys. Rev. D{\bf 47} (1993), 2918. 
 %
\bibitem{ECQ}
C. A. de S. Pires and O. P. Ravinez, Phys. Rev. D {\bf 58}(1998), 35008;
C. A. de S. Pires, Phys. Rev. D {\bf 60}(1999), 075013.
\bibitem{DCS}
F.  Cuypers, S. Davidson, Eur. Phys. J.  C{\bf 2} (1998), 503,
G. Tavares-Velasco, J. J. Toscano, Phys. Rev.  D{\bf 65} (2002), 013005.
\bibitem{typeIISS331}
J. C. Montero, C. A. de S. Pires, V. Pleitez,  Phys. Lett. B{\bf 502} (2001), 167; 
M. B. Tully, G. C. Joshi, Phys. Rev. D{\bf 64} (2001), 011301.
\bibitem{neutrinos331}
Y. Okamoto, M. Yasue, Phys. Lett. B{\bf 466} (1999), 267;
T. Kitabayashi, M. Yasue, Nucl. Phys. B{\bf 609} (2001), 61;
J. C. Montero, C. A. de S. Pires, V. Pleitez, Phys. Rev. D{\bf 66} (2002), 113003. 
\bibitem{landaupole}
A. G. Dias, R. Martinez, V. Pleitez,  Eur. Phys. J. C{\bf 39}(2005), 101. See also,  A. G. Dias, V. Pleitez, Phys. Rev. D{\bf 80}(2009), 056007.
\bibitem{trulyminimal}
F.  Queiroz, C. A. de S. Pires, P. S. Rodrigues da Silva, Phys. Rev. D{\bf 82} (2010), 065018.
\bibitem{combinetypeI-II}
W.  Caetano,  D. Cogollo, C. A. de S. Pires, P. S. Rodrigues da Silva,  Phys. Rev. D{ \bf 86}  (2012), 055021.
\bibitem{twotriplets}
J. G. Ferreira, Jr,  P. R. D. Pinheiro, C. A. de S. Pires, P. S. Rodrigues da Silva,  Phys. Rev. D{\bf 84} (2011), 095019.
\bibitem{rho}
In this point we would like to call the attention to a difficulty concerning the reduced 3-3-1 model. It was recently showed that this version  faces difficulty to accommodate the current experimental value of the parameter $\rho$, as discussed in :
P.V. Dong, D. T. Si, Phys. Rev. D{\bf 90} (2014) 11, 117703.

\bibitem{typeI}
M. Gell-Mann, P. Ramond, and R. Slansky, in {\it supergravity}, edited by P. van Nieuwenhuizen and D. Z. Freedman (North-Holland, amstrdam, 1979);
T. Yanagida, in {\it proceedings of the Workshop on the Unified Theory and the Baryon number  in the Universe}, edited by O. Sawada and A. Sugamoto (KEK Report No. 79-18, Tsukuba, Japan, 1979);
R. N. Mohapatra  and G. Senjanovic, Phys. Rev. Lett. {\bf 44},(1980), 912.
\bibitem{typeII}
M. Magg, C. Wetterich, Phys. Lett. B{\bf 94} (1980), 61;
R. N. Mohapatra, G. Senjanovic, Phys. Rev. D{\bf 23}(1981), 165; 
E. Ma, U. Sarkar, Phys. Rev. Lett. {\bf 80}(1998), 5716 .
\bibitem{grimus}
W. Grimus, L. Lavoura, B. Radovcic, Phys. Lett. B{\bf 674} (2009), 117.
\bibitem{pal}
 P. B. Pal, Phys. Rev. D{\bf 52} (1995), 1659;
 A. G. Dias, C. A. de S. Pires, P. S. Rodrigues da Silva, Phys. Rev. D{\bf 68} (2003), 115009. 
\bibitem{naturallylightRH}
Alex G. Dias (Sao Paulo U.), C.A. de S.Pires, P.S. Rodrigues da Silva,  Phys. Lett. B{\bf 628} (2005), 85. 
\bibitem{LRversion}
This scheme also works for the left-right extension of this model, see Ref.: A.  G. Dias, C. A. de S. Pires, P . S. Rodrigues da Silva, Phys. Rev. D{\bf 82} (2010), 035013. 
\bibitem{realization}
D. Cogollo, H. Diniz, C. A. de S.Pires, Phys. Lett. B{\bf 677} (2009), 338;
N. A. Ky, N. T. H. Van, Phys. Rev. D{\bf 72}(2005), 115017; A. Palcu, Mod. Phys. Lett. A{\bf 21}(2006), 2591; P. V. Dong, H. N. Long, Phys. Rev. D{\bf 77}(2008), 057302; D. Cogollo, H. Diniz, C. A. de S.Pires, P.S. Rodrigues da Silva, Eur. Phys. J. C{\bf 58}(2008), 455.
\bibitem{dodelson}
KeV sterile neutrino as warm dark matter candidate was first proposed by S. Dodelson, L. M. Widrow, Phys. Rev. Lett.{\bf 72} (1994), 17;
X-D. Shi, G. M. Fuller, Phys. Rev. Lett. {\bf 82} (1999), 2832.
\bibitem{ISSgeneral}
R.  N. Mohapatra,	Phys. Rev. Lett. {\bf 56}, 561  (1986); 
R. N. Mohapatra, J. W. F. Valle, Phys. Rev. D{\bf 34} , 1642 (1986).
A. G. Dias,  C. A. de S. Pires, P. S. Rodrigues da Silva,  Phys. Rev.  {\bf D 84}, 053011 (2011).
\bibitem{ISS331}
A. G. Dias, C. A. de S. Pires, P. S. Rodrigues da Silva, A. Sampieri, Phys. Rev. D{\bf 86} (2012), 035007.
\bibitem{moreonISS331}
To see more on ISS mechanism in 3-3-1 model, see: 
M. E. Catao, R. Martnez, F. Ochoa, Phys. Rev. D {\bf 86} (2012),073015;
A. Palcu , arXiv:1408.6518.
\bibitem{DM331}
C. A. de S. Pires, P. S. Rodrigues da Silva,  JCAP {\bf 0712} (2007), 012 ;
J. K. Mizukoshi, C. A. de S. Pires, F. S. Queiroz, P. S.  Rodrigues da Silva, Phys. Rev. D{\bf 83} (2011), 065024.
\bibitem{Higgspheno}
A. Alves, E. Ramirez Barreto, A. G. Dias, C.A. de S. Pires, F. S. Queiroz, P. S. Rodrigues da Silva, Phys. Rev. D{\bf 84} (2011), 115004 ;
J. D. Ruiz-Alvarez, C. A. de S.Pires, Farinaldo S. Queiroz, D. Restrepo, P. S. Rodrigues da Silva, Phys. Rev. D{\bf 86} (2012), 075011 ;
A. Alves, E. Ramirez Barreto, A. G. Dias, C. A. de Pires, Farinaldo S. Queiroz, P. S. Rodrigues da Silva, Eur. Phys. J. C{\bf 73} (2013) 2, 2288 ;
W. Caetano, C. A. de S. Pires, P. S. Rodrigues da Silva, D. Cogollo, Farinaldo S. Queiroz, Eur. Phys. J. C{\bf 73} (2013) 10, 2607.
\bibitem{zprime}
Y. A. Coutinho, V. Salustino Guimares, A. A. Nepomuceno,
Phys. Rev. D {\bf 87} (2013), 115014; E.
Ramirez Barreto, Y. A. Coutinho, J. Sa Borges, Phys. Lett. B
{\bf 689} (2010), 36.
\bibitem{diego}
D. Cogollo, Alma X. Gonzalez-Morales, Farinaldo S. Queiroz, Patricia R. Teles ,  JCAP {\bf 1411} (2014) 11, 002 .

\end{thebibliography}
\end{document}